# Two-Staged Magnetoresistance Driven by Ising-like Spin Sublattice in $SrCo_6O_{11}$

S. Ishiwata[1,2*], I. Terasaki[2], F. Ishii[3], N. Nagaosa[3,4], H. Mukuda[5], Y. Kitaoka[5], T. Saito[1], and M. Takano[1]

[1]*Institute for Chemical Research (ICR), Kyoto University, Uji, Kyoto 611-0011, Japan.*

[2]*Department of Applied Physics, Waseda University, 3-4-1 Okubo, Shinjuku, Tokyo 169-8555, Japan.*

[3]*Tokura Spin SuperStructure Project (SSS), ERATO, Japan Science and Technology Agency, AIST Tsukuba Central 4, 1-1-1 Higashi, Tsukuba, Ibaraki 305-8562, Japan.*

[4]*CREST, Department of Applied Physics, University of Tokyo, 7-3-1, Hongo, Bunkyo-ku, Tokyo 113-8656, Japan.*

[5]*Department of Materials Engineering Science, Osaka University, 1-3 Machikaneyama-cho, Toyonaka, Osaka 560-8531, Japan.*

**A two-staged, uniaxial magnetoresistive effect has been discovered in $SrCo_6O_{11}$ having a layered hexagonal structure. Conduction electrons and localized Ising spins are in different sublattices but their interpenetration makes the conduction electrons sensitively pick up the stepwise field-dependence of magnetization. The stepwise field-dependence suggests two competitive interlayer interactions between ferromagnetic Ising-spin layers, i.e., a ferromagnetic nearest-layer interaction and an antiferromagnetic next-nearest-layer interaction. This oxide offers a unique opportunity to study nontrivial interplay between conduction electrons and Ising spins, the coupling of which can be finely controlled by a magnetic field of a few Tesla.**

Magneto-transport phenomena are one of the most important issues in current condensed matter physics. The energy scale of magnetic field is usually very small compared with that of electron-electron interactions. Namely, the former is ~1 meV even at 10 Tesla, whereas the latter is typically 1 eV. Therefore, it is surprising and important that electronic states prove to be sensitively field-dependent in some materials. One of such phenomena is known as the colossal magnetoresistance (CMR) in perovskite-type manganites close to a multicritical point [1-6]. The competition between the ferromagnetic metallic state due to the double-exchange interaction and the charge/orbital-ordered insulating state provides a subtle, balanced sensitivity to external parameters such as magnetic field, pressure, and chemical composition. Also having been studied extensively are the giant magnetoresistance (GMR) and spin-valve effect taking place at low fields of ~mT in metallic thin films having artificially constructed magnetic infrastructures [7].

In this paper, we present a novel type of GMR system, which is $SrCo_6O_{11}$ crystallizing in a layered hexagonal structure. The most important features of this system are as follows. (i) The conduction electrons and the localized moments in different, interpenetrating sublattices are coupled with each other. (ii) The localized moments are Ising-like spins with a strong uniaxial anisotropy along the $c$ axis. (iii) The field dependence of these properties along the $c$ axis is two-staged, reflecting the presence of such competitive interactions in the Ising spin system as leading to stepwise changes of the ordered superstructure under a magnetic field of a few Tesla.

Single crystals with a typical size of $0.25 \times 0.25 \times 0.05$ mm$^3$ were prepared under high pressure and high temperature as reported elsewhere [8]. This oxide crystallizes in the $R$-type hexagonal ferrite structure (space group; $P6_3/mmc$, $a = 5.609(4)$ Å, $c = 12.58(1)$ Å, $V = 342.7(5)$ Å$^3$, $Z = 2$) composed of infinitely stacked Co-O layers of the Kagomé type and two types of intervening Co-O pillars (See the inset of Fig. 1(a)): The edge-sharing $Co(1)O_6$ octahedra form the $Co(1)_3O_8$ Kagomé planes, while the $Co(2)_2O_9$ dimerized octahedra and the $Co(3)O_5$ trigonal bipyramids, which are both arranged in hexagonal symmetry, work as pillars.

Magnetization was measured with a MPMS-XL (Quantum Design). Resistivity (for contact configurations, see the insets of Figs. 1(b) and 1(c)) was measured with a PPMS (Quantum Design). Spin-echo NMR measurements were performed using a conventional phase-coherent-type spectrometer. A first-principles electronic structure calculation in the scalar relativistic regime was performed with the all-electron full-potential linear augmented plane wave (FLAPW) method within local spin density approximation (LSDA) [9]. The improved tetrahedron method [10] was used for the Brillouin-zone integration with the k mesh (12,12,6) resulting in 76 irreducible k points during the self-consistent field and density of states (DOS) calculation. Muffin-tin sphere radii were set to be 1.0, 1.0, and 0.6 Å for Sr, Co, and O, respectively. Plane wave cutoffs were 15 and 60 Ry for the LAPW basis and the star function, respectively.

First we show a remarkable interplay between the conduction electrons and the localized moments in the magnetically ordered states. Figure 1(a) shows the magnetization ($M$) plotted against external magnetic field ($H$). For the magnetization along the $c$ axis at 5 K, a well-defined $M_c/M_c^s = 1/3$ plateau extends from ~0 to 2.5 T, above which it jumps to another $M_c/M_c^s = 3/3$ plateau ($M_c^s \sim 4$ $\mu_B$ per formula unit). This makes a remarkable contrast with the small in-plane magnetization ($M_{ab}$), suggesting that the local moment has an Ising-like anisotropy. The plateaus become somewhat smeared with increasing temperature and eventually fade away above 20 K. Most remarkably, the out-of-plane resistivity, $\rho_c$, changed concomitantly at exactly the same fields as shown in Fig. 2(b). Strong electromagnetic coupling along the $c$ axis is manifested by the large magnetoresistance (MR), $(\rho_c(H)-\rho_c(0))/\rho_c(0)$, at 5 K which is negative and as large as -19 % at 2 T and -56 % at 5 T. The resistivity plateaus tend to smear at higher temperatures as the magnetization plateaus do. The in-plane MR is also negative and considerably field-dependent, but the behavior is bumpy rather than plateau-like (Fig. 2(c)). These bumps suggest critical scattering near the field-induced magnetic transitions. Concerning the magnitude of MR, the in-plane MR at 5 K and 5 T is considerably smaller than the out-of-plane MR. The temperature dependence is opposite: The in-plane MR at 5 T decreases as temperature is decreased from 15 K to 5 K, suggesting the importance of thermal factor. Around 0 T both the resistivity and magnetization measured at 5 K show a small hysteresis, which further evidences the electromagnetic coupling. The experimental results described above indicate that the ordered structure of the localized Ising spins changes stepwise depending upon external field and that the electromagnetic coupling is accordingly anisotropic as seen in the difference in magnitude and also the opposite temperature dependence of MR.

Next, we studied the behavior at higher temperatures including the transition region. Figure 2(a) shows $M_c/H$ plotted against temperature measured at 0.01 and 1 T. Curie-Weiss-like behavior at high temperatures is followed by anomalies taking place below 20 K. The data for 1 T between 150 and 300 K were fit by the Curie-Weiss law with a temperature independent term included, $\chi(T) = \chi_0 + C / (T - \theta)$, $\chi_0 = 5.7(1) \times 10^{-3}$ emu / Oe mol, $C$ = 3.14(4) emu K / Oe mol and $\theta$ = 32(1) K, indicating that a ferromagnetic interaction is dominant. An abrupt upturn after a small drop was observed below 20 K ($T_{c1}$) for $H$ = 0.01 T. Then, the temperature dependence of $M_c/H$ became hysteretic below 12 K ($T_{c2}$). As seen in Fig. 2(b), the resistivity is metallic in both directions but anisotropic as $\rho_c/\rho_{ab}$ ~ 20 [11]. Accordingly, Co(1)$_3$O$_8$ Kagomé planes should be more conductive than the pillars. Both $\rho_{ab}$ and $\rho_c$ decrease remarkably below $T_{c1}$ and show another change of slope at $T_{c2}$. Thus the magnetic transition seems to proceed in two steps, the one at $T_{c1}$ having a more pronounced influence upon resistivity. We note here that the negative MR (5 T) is perceivable up to 50 K, implying short-ranged magnetic correlations being developed far above $T_{c1}$. The non-linear field dependence of $M_c$ at 25 K seen in Fig. 1(a) also supports this picture.

It is known through our NMR measurements that the Co(1) and Co(2) atoms are almost nonmagnetic, while the Co(3) atoms carry localized spins [12]. Moreover, neutron diffraction studies have revealed that the magnetic structure in the 1/3-plateau state is such that the spins are ferromagnetically oriented in every Co(3)-layer normal to the $c$ axis and that the layer moment changes its direction along the $c$-axis with a periodicity of 3$c$ as -↑-↑-↓-↑-↑-↓- [13]. The NMR spectra in this state and the saturated 3/3-plateau state are compared in Fig. 3. In the 3/3 plateau state, both the Co(1) and Co(2) components are domed because their quadrupole splittings are smeared out by magnetic perturbations arising from supertransferred hyperfine interactions with the Co(3) atoms (Fig.3 (a)). In the 1/3-plateau state, the Co(1)-component shows a remarkable change, such that the component is split into two. The blue-colored component (Co(1)$^M$), which has a common shape in both the 1/3 and 3/3 states, comes from the Co(1) ions sandwiched by a pair of Co(3, ↑) layers. The yellow-colored component (Co(1)$^{NM}$) coming from the Co(1) ions sandwiched by Co(3, ↑) and Co(3, ↓) layers has a well-articulated shape because the supertransferred hyperfine interactions smearing out the quadrupole splitting are cancelled by centrosymmetry at Co(3).

The first-principle calculations have successfully reproduced the magnetic moment at the Co(3) site. When the spin configurations for each site are assumed to be ferromagnetic, an antiferromagnetic state (Co(1) ↓ Co(2) ↓ Co(3) ↑) is obtained as the valid solution. The magnetic moment (inside the muffin-tin sphere) is calculated to be 1.5 $\mu_B$ for Co(3), which is much larger than those for octahedrally coordinated Co(1) and Co(2) (< 0.3 $\mu_B$). Thus, it is the Co(3) atoms that play the major role in the magnetism, whereas Co(1) and Co(2) atoms are expected to be Pauli paramagnetic. The partial density of states (PDOS) projected on partial states within the Co muffin-tin spheres for the antiferromagnetic state is plotted for Co(1)/Co(2) and Co(3) in Figs. 4(a) and 4(b), respectively. A half metallic character with a gap of 0.1 eV is clearly seen near the Fermi level, where a small density of holes exists in the $t_{2g}$ bands. The gap is formed by a crystal field splitting. However, in fact, the presence/absence of the gap is a subtle issue. Nevertheless, we expect that a "pseudogap" near the Fermi level will remain and lead to the enhanced MR.

The alternate stacking of Co(2)-pillared Kagomé layers and Co(3)-magnetic layers builds up a novel type of spin-valve system. Here we propose a kind of Kondo lattice model for the present compound. The conduction within the Co(2)-pillared Kagomé system is biased by the localized Co(3) spins so that the density of states of the conduction electrons is almost zero for one of the spins. In this context, we assumed that the in-plane Co(3) spin arrangement is ferromagnetic below $T_{c1}$, and that the three-fold periodicity along the $c$ axis develops below $T_{c2}$. When the out-of-plane spin arrangement of Co(3) changes as follows: the ground state, -↑-↕-↓- ($M_c/M_c^s = 0$) → the ferrimagnetic state, -↑-↑-↓- ($M_c/M_c^s = 1/3$) → the ferromagnetic state, -↑-↑-↑- ($M_c/M_c^s = 3/3$), the out-of-plane resistivity should decrease in two steps. The ground state has no spontaneous moment presumably because two out of three Co(3) layers couple antiferromagnetically and the remaining one is random. A similar magnetic order is realized in CeSb which has been studied in the framework of the axial next-nearest neighbor Ising (ANNNI) model [14]. In this model, Ising spins are coupled ferromagnetically in every layer, while along the out-of-plane direction ferromagnetic interactions between the nearest layers and antiferromagnetic interactions between the next nearest layers are competitive to yield a partially disordered state [15]. In the case of $SrCo_6O_{11}$ conduction-electron-mediated ferromagnetic interactions between the nearest layers seem to be competitive with antiferromagnetic next nearest-layer superexchange interactions.

In summary, we found a novel magneto-transport phenomenon in $SrCo_6O_{11}$. This material can be regarded as a two-staged spin-valve system where the spin-polarized conduction electrons couple with the Ising-like spins. Although the origin of the threefold-superperiodic spin structure is an open question, "two-staged coupling" in this material offers a unique opportunity to study nontrivial interplay between conduction electrons and Ising spins.

**ACKNOWLEDGEMENTS**

We thank D. I. Khomskii, S. Miyashita, Y. Shimakawa and H. Kageyama for fruitful discussion, and M. Azuma and W. Kobayashi for technical support. This work was supported by Grants-in-Aid for Scientific Research A14204070 from the MEXT, that for COE Research on Elements Science, that for 21$^{st}$ Century COE Programs at Kyoto Alliance for Chemistry, and that for 21$^{st}$ Century COE Programs at Waseda University for Physics.  S. I. is a research fellow of JSPS.

**FIG. 1: (Color online) (a) Out-of-plane magnetization curves ($H // c$) at 5, 15, and 25 K. In-plane magnetization ($H // a$) at 5 K, represented by the broken line, is also appended. (b) Normalized out-of-plane resistivity, $\rho_c(H)/\rho_c(0)$, and (c) normalized in-plane resistivity, $\rho_{ab}(H) / \rho_{ab}(0)$, at 5, 15, and 25 K under a magnetic field applied parallel to the $c$ axis. For clarity, the resistivity curves at 5 K and 15 K have been shifted downward. In order to indicate the hysteresis, the resistivity curves obtained on reducing magnetic fields are denoted with the dashed curves, and the sequential numbers with arrows showing the order of measurements are appended. Schematic crystal structure and sample configurations for the resistivity measurements were illustrated as insets.**

**FIG. 2: (Color online) Temperature dependence of the magnetic and electronic properties. (a) The DC magnetization along the $c$ axis divided by a field, $M/H$,**

obtained on field-heating after zero-field-cooling (ZFC) and field-cooling (FC) in a magnetic field of 0.01 T and 1 T. (b) The in-plane and the out-of-plane resistivities, $\rho_{ab}$ and $\rho_c$, respectively, with or without the magnetic field of 5 T along the *c* axis.

FIG. 3: (Color online) Co-NMR spectra measured at 4.2 K with a frequency of 92.3 MHz for the 3/3-plateau state (a) and with a frequency of 24.4 MHz for the 1/3-plateau state (b). The blue area represents a spectral component featured by a large shift caused by the presence of a transferred internal magnetic field, while the yellow component with a small shift is affected by the nuclear quadrupole interactions only. The insets are schematic illustrations of the corresponding magnetic states.

FIG. 4: (Color online) Calculated PDOS per Co muffin-tin sphere for Co(1), Co(2) (a) and Co(3) (b) in the antiferromagnetic solution (Co(1) ↓ Co(2) ↓ Co(3) ↑). (c) Schematic illustration of a spin-valve effect in the 1/3-plateau state. The thick arrows on the pillars indicate directions of the Co(3) moments, while the small circled arrows denote spin polarization of the conduction electrons. The scattering rate along the out-of-plane is enhanced when the moment of Co(3) is antiparallel to that of the conduction electrons.

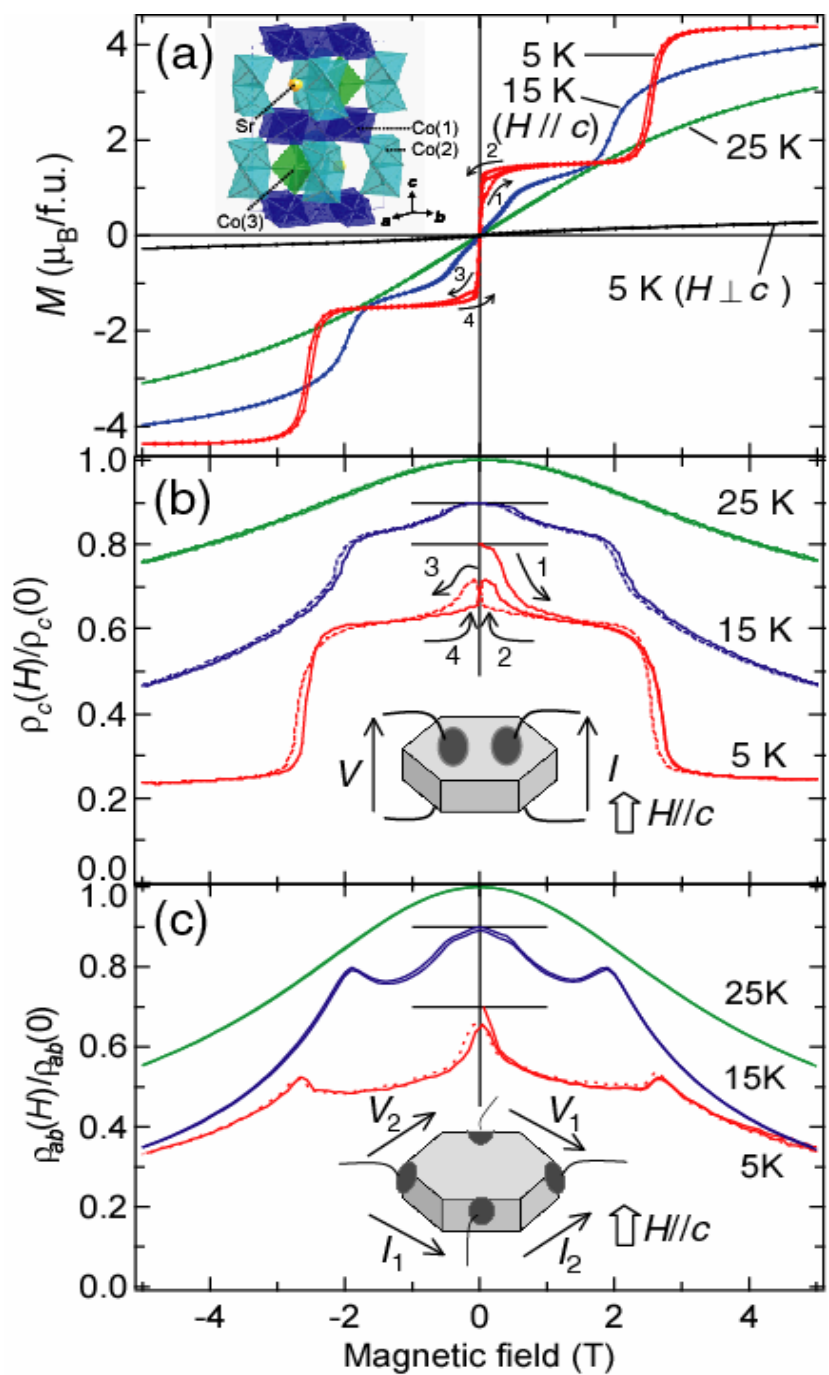

Fig. 1

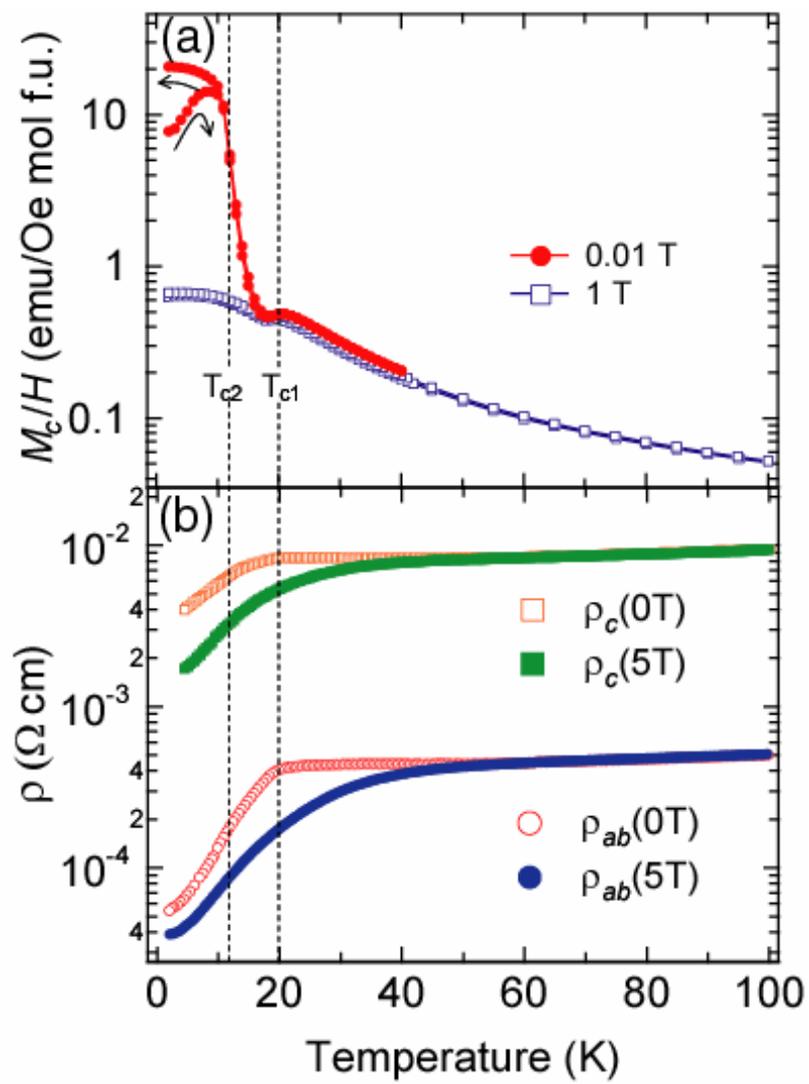

Fig. 2

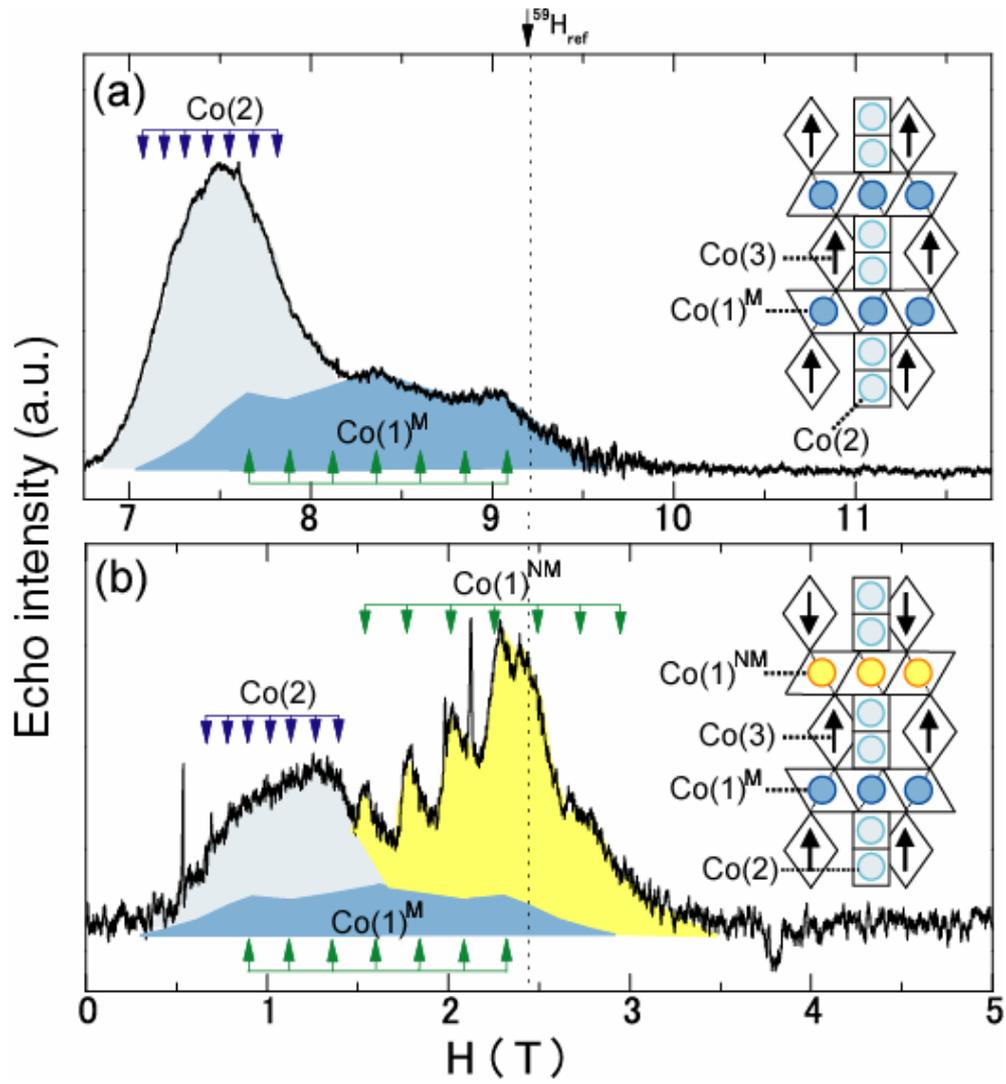

Fig. 3

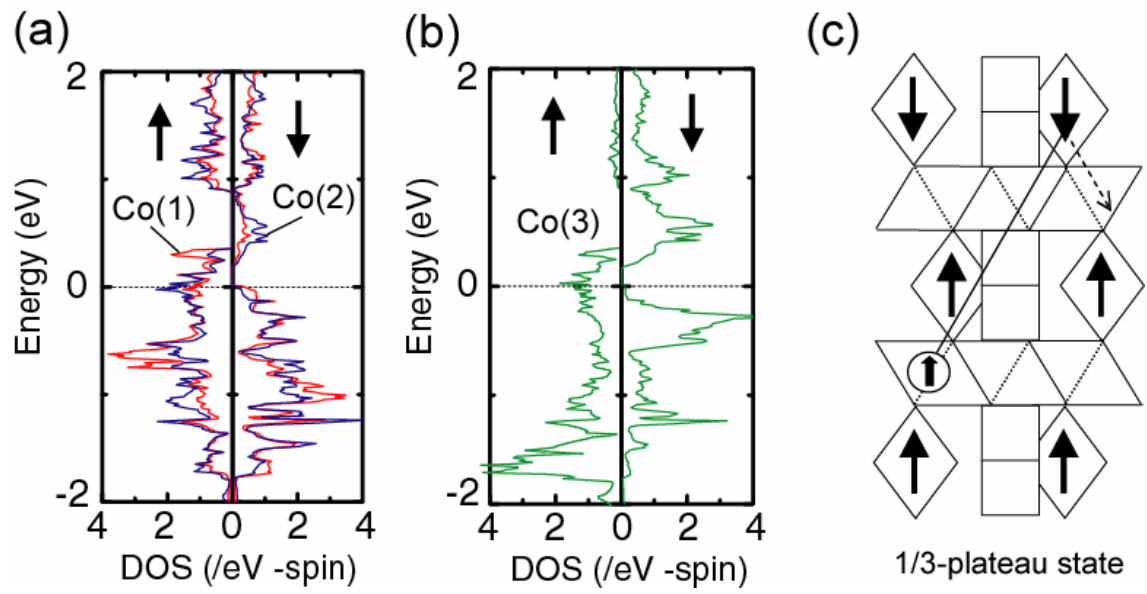

Fig. 4